\documentclass[preprint,12pt]{elsarticle}
\usepackage{bbding}
\usepackage{tipa}
\usepackage{pifont}
\usepackage{amsfonts}
\usepackage{mathrsfs}

\usepackage[english]{babel}
\usepackage{pgf,pgfarrows,pgfnodes,pgfautomata,pgfheaps}
\usepackage[space]{ctex}
\usepackage{amsmath}
\usepackage[latin1]{inputenc}
\usepackage{dsfont}
\usepackage{multicol}
\usepackage{amssymb}
\usepackage{amsthm}
\journal{discrete mathematics}

\newtheorem{theorem}{Theorem}[section]
\newtheorem{definition}[theorem]{Definition}
\newtheorem{lemma}[theorem]{Lemma}

\newtheorem{corollary}[theorem]{Corollary}
\newtheorem{example}[theorem]{Example}

\begin{document}
\begin{frontmatter}



\title{Bounds and Constructions for $\overline{3}$-Strongly Separable Codes with Length $3$}

\author[label1]{Xuli Zhang}
\author[label2]{Jing Jiang}
\author[label2]{Minquan Cheng}
\address[label1]{School of Mathematics and Statistics, Guangxi Normal University}
\address[label2]{Guangxi Key Lab of Multi-source Information Mining \& Security, Guangxi Normal University}

\begin{abstract}
As separable code (SC, IEEE Trans Inf Theory 57:4843-4851, 2011) and frameproof code (FPC, IEEE Trans Inf Theory 44:1897-1905, 1998) do in multimedia fingerprinting,
strongly separable code (SSC, Des. Codes and Cryptogr.79:303-318, 2016) can be also used to construct anti-collusion codes. Furthermore, SSC is better than FPC and SC in the applications for multimedia fingerprinting since SSC has lower tracing complexity than that of SC (the same complexity as FPC) and weaker structure than that of FPC. In this paper, we first derive several upper bounds on the number of codewords of $\overline{t}$-SSC.
Then we focus on $\overline{3}$-SSC with codeword length $3$, and obtain the following two main results:
(1) An equivalence between an SSC and an SC.
(2) An improved lower bound $\Omega (q^{5/3}+q^{4/3}-q)$ on the size of a $q$-ary SSC when $q=q_1^6$ for any prime power $q_1\equiv\ 1 \pmod 6$, better than the before known bound $\lfloor\sqrt{q}\rfloor^{3}$, which is obtained by means of difference matrix and the known result on the subset of $\mathbb{F}^{n}_q$ containing no three points on a line.
\end{abstract}

\begin{keyword}
Multimedia fingerprinting, separable code, strongly separable code, forbidden configuration, difference matrix.
\end{keyword}
\end{frontmatter}

\section{Introduction}
\label{pre}
Let $n$, $M$ and $q$ be positive integers, and $Q$ an alphabet with $|Q|=q$.
A set $\mathcal{C} =\{{\bf c}_1,{\bf c}_2,\ldots, {\bf c}_M\} \subseteq Q^n$ is called an $(n,M,q)$ code
and each ${\bf c}_i$ is called a codeword.
Without loss of generality, we may assume $Q=\{0,1,\ldots,q-1\}$.
When $Q=\{0,1\}$, we also use the word \lq\lq binary\rq\rq.

For any code $\mathcal{C} \subseteq Q^n$, we define the set of $i$th coordinates of $\mathcal{C}$ as
$$\mathcal{C}(i) =\{ {\bf c}(i) \in Q  \ | \ {\bf c}=({\bf c}(1), {\bf c}(2), \ldots, {\bf c}(n))^{T} \in \mathcal{C}\}$$
for any $1 \le i \le n$. For any subset of codewords $\mathcal{C}_0\subseteq \mathcal{C}$, we
define the descendant code of $\mathcal{C}_0$ by
$$
{\sf desc}(\mathcal{C}_0) = \{({\bf x}(1),{\bf x}(2),\ldots,{\bf x}(n))^{T}  \in Q^n \ |\ {\bf x}(i) \in \mathcal{C}_0(i), 1 \le i \le n\},
$$
that is, $${\sf desc}(\mathcal{C}_0)=\mathcal{C}_{0}(1)\times \mathcal{C}_{0}(2)\times \ldots \times  \mathcal{C}_{0}(n).$$
Clearly the set ${\sf desc}(\mathcal{C}_0)$ consists of the $n$-tuples that could be
produced by a coalition holding the codewords in $\mathcal{C}_0$.
\begin{definition}\rm(\cite{C.M,JCM})
\label{de1}
Let $\mathcal{C}$ be an $(n,M,q)$ code and $t \geq 2$ be an integer.
\begin{itemize}
\item $\mathcal{C}$ is a $\overline{t}$-separable code, or $\overline{t}$-SC$(n,M,q)$, if for any $\mathcal{C}_{1}$, $\mathcal{C}_{2} \subseteq \mathcal{C}$ such that
 $1 \leq |\mathcal{C}_{1}| \leq t$,
 $1\leq |\mathcal{C}_{2}| \leq t$,
 and  $\mathcal{C}_{1} \neq \mathcal{C}_{2}$,
 we have ${\sf desc}(\mathcal{C}_{1}) \neq\ {\sf desc}(\mathcal{C}_{2})$,
 that is there is at least one coordinate $i$, $1\leq i\leq n$, such that $\mathcal{C}_{1}(i)\neq \mathcal{C}_{2}(i)$.
\item $\mathcal{C}$ is a  strongly $\overline{t}$-separable code, or $\overline{t}$-SSC$(n,M,q)$, if for any $\mathcal{C}_{0}\subseteq \mathcal{C}$ such that $1\leq|\mathcal{C}_{0}|\leq t$, we have $\cap_{\mathcal{C}' \in \hbox{S}(\mathcal{C}_{0})}\mathcal{C}'= \mathcal{C}_{0}$,
 where $S(\mathcal{C}_{0})=\{\mathcal{C}'\subseteq\mathcal{C}| {\sf desc}(\mathcal{C}') = {\sf desc}(\mathcal{C}_{0})\}$.
\item $\mathcal{C}$ is a $t$-frameproof code, or $t$-FPC$(n,M,q)$,
if for any $\mathcal{C}' \subseteq \mathcal{C}$ such that $|\mathcal{C}'| \le t$,
it holds that ${\sf desc}(\mathcal{C}') \bigcap \mathcal{C} = \mathcal{C}'$, that is,
for any ${\bf c} =({\bf c}(1),\ldots, {\bf c}(n))^{T}\in \mathcal{C} \setminus \mathcal{C}'$,
there is at least one coordinate $i$, $1 \le i \le n$, such that ${\bf c}(i) \not\in \mathcal{C}'(i)$.
\end{itemize}
\end{definition}

Separable code, strongly separable code and frameproof code can be used to construct anti-collusion codes which
can effectively trace and even identify the sources of pirate copies of copyrighted multimedia data. However the anti-collusion codes constructed by $\overline{t}$-separable codes have tracing complexity $O(M^{t})$, and those constructed by strongly  $\overline{t}$-separable codes and $t$-frameproof codes have tracing complexity $O(M)$. Compared with frameproof codes, strongly separable codes have an advantage in copyright protection since a strongly $\overline{t}$-separable code has the same traceability as a $t$-frameproof code but has more codewords than a $t$-frameproof code, see \cite{JCM}. Strongly separable codes can be also used to study the classic digital fingerprinting codes such as identifiable parent property (IPP) codes \cite{HLLT}, frameproof codes \cite{B,S.S.W}, perfect hash families (PHFs) \cite{B4,S.W.C} and so on. In this paper, we will pay our attention to strongly separable codes.

Since the parameter $M$ of a $\overline{t}$-SSC$(n,M,q)$ corresponds to the number of fingerprints
assigned to authorized users who purchased the right to access the copyrighted multimedia data,
we should try to construct strongly separable codes with $M$ as large as possible, given length $n$.
Let $M(\overline{t},n,q) = \hbox{max} \{M \ | \ \hbox{there exists a } \overline{t} \hbox{-SSC}(n,M,q)\}$.
A $\overline{t}$-SSC$(n,M,q)$ is said to be optimal if $M = M(\overline{t},n,q)$. Similarly, a $\overline{t}$-SC$(n,M,q)$ (or a $t$-FPC$(n,M,q)$) is optimal if $M$ is the largest possible value given $n$, $q$ and $t$.

According to the relationship between strongly separable codes and separable codes, $\overline{t}$-SSC$(n,M,q)$
with parameters $(t,n)=(2,2)$,
and $(2,3)$ were discussed in \cite{JCM}. When $t\geq 3$,
the structure of $\overline{t}$-SSCs becomes more complex so that little is known about $\overline{t}$-SSCs.
The remainder of the paper is organized as follows. In Section \ref{se-upper bounds},
according to the results on separable codes and frameproof codes, we first derive several upper bounds on $M(\overline{t},n,q)$.
And then by showing an equivalence between a $\overline{3}$-SC$(3,M,q)$ and a $\overline{3}$-SSC$(3,M,q)$,
a tighter upper bound ($M(\overline{3},3,q)\leq \frac{3}{4}q^2$) and a lower bound ($M(\overline{3},3,q)\geq \lfloor\sqrt{q}\rfloor^3$) are derived.
In Section \ref{se-construction}, a construction for $\overline{3}$-SSC$(3,M,q)$ will be then provided by means of difference matrix and the known result on the subset of $\mathbb{F}^{n}_q$ containing no three points on a line.
As a consequence, a new lower bound $\Omega (q^{5/3}+q^{4/3}-q)$ on $M(\overline{3},3,q)$ is derived.

\section{Upper bounds}
\label{se-upper bounds}
In this section, we first investigate the relationships among SC, SSC and FPC, and then derive the upper bounds on $M(\overline{t},n,q)$  according to the relationships.
\subsection{SC, SSC and FPC}\

The relationship between SC and FPC was proposed in \cite{C.M}.
\begin{lemma} \rm (\cite{C.M})
\label{le1}
Any $t$-FPC$(n, M, q)$ is a $\overline{t}$-SC$(n, M, q)$, $t \geq 1$. Conversely any $\overline{t}$-SC$(n, M, q)$ is a $(t-1)$-FPC$(n, M, q)$, $t \geq 2$.
\end{lemma}
Jiang et al. \cite{JCM} established the following relationships among SC, SSC and FPC.
\begin{lemma}\rm(\cite{JCM})
\label{le2}
Any $t$-FPC$(n, M, q)$ is a $\overline{t}$-SSC$(n, M, q)$.
\end{lemma}
The following example shows that the converse of Lemma \ref{le2} does not always hold.
\begin{example}\rm (\cite{JCM}) The following $(3, 4, 2)$ code $\mathcal{C}$ is a $\overline{2}$-SSC$(3,4,2)$, but is not a $2$-FPC$(3,4,2)$.
\begin{eqnarray*}
\begin{array}{c}
  \ \ \ \ \  {\bf c}_1 \  {\bf c}_2 \  {\bf c}_3 \  {\bf c}_4 \\
\mathcal{C}=
\left(\begin{array}{cccc}
 0 \ & 1 \ & 0 \ & 0 \ \\
 0 \ & 0 \ & 1 \ & 0  \ \\
 0 \ & 0 \ & 0 \ & 1  \
  \end{array}\right)
  \end{array}
\end{eqnarray*}
\end{example}

\begin{lemma} \rm (\cite{JCM})
\label{le3}
Any $\overline{t}$-SSC$(n,M,q)$ is a $\overline{t}$-SC$(n,M,q)$.
\end{lemma}

Although, the converse of Lemma \ref{le3} does not always hold, when $t=n=2$, Jiang et al. proved that the converse of Lemma \ref{le3} is also true. 

\begin{example} \rm (\cite{JCM})
Let ${\bf c}_{i}$, $1\leq i\leq5$, be the $i$th codeword of the following code $\mathcal{C}$, then $\mathcal{C}$ is a $\overline{2}$-SC$(3,5,2)$, but not a $\overline{2}$-SSC$(3,5,2)$, because ${\sf desc}(\{{\bf c}_{1},{\bf c}_{5})=$\ ${\sf desc}(\{{\bf c}_{2},{\bf c}_{3},{\bf c}_{4}\})$.
\begin{eqnarray*}
\mathcal{C}=\left(
\begin{array}{ccccc}
0&1&0&0&1\\
0&0&1&0&1\\
0&0&0&1&1
\end{array}
\right) \ \ \
\end{eqnarray*}
\end{example}

\begin{lemma} \rm (\cite{JCM})
\label{le3.1}
A $(2,M,q)$ code $\mathcal{C}$ is a $\overline{2}$-SSC$(2,M,q)$ if and only if $\mathcal{C}$ is a $\overline{2}$-SC$(2,M,q)$.
\end{lemma}

\begin{example} \rm
The following code $\mathcal{C}$ is an optimal $\overline{3}$-SC$(3,3,2)$, and we can check that it is also a $\overline{3}$-SSC$(3,3,2)$.
\begin{eqnarray*}
\mathcal{C}=\left(
\begin{array}{ccc}
0&1&1\\
0&1&0\\
0&0&1
\end{array}
\right) \ \ \
\end{eqnarray*}
\end{example}

Furthermore, it is very interesting that the converse of Lemma \ref{le3} also holds for $t=n=3$ and $q \geq 3$. We first state the two fuseful results. From Lemmas \ref{le1} and \ref{le3}, the following statement holds.
\begin{corollary} \rm
\label{co1}
Any $\overline{t}$-SSC$(n,M,q)$ is a $(t-1)$-FPC$(n,M,q)$ where $t\geq 2$.
\end{corollary}
\begin{lemma} \rm
\label{le4}
Suppose $\mathcal{C}$ is a $\overline{3}$-SC$(3,M,q)$. Then for any
$\mathcal{C}_{0} \subseteq \mathcal{C}$ with $ |\mathcal{C}_{0}| \leq 3$,
and any ${\bf c} \in \mathcal{C}_{0}$, the Hamming distance $d({\bf c}, {\bf c}') \geq 2$ holds for any
${\bf c}' \in {\sf desc}(\mathcal{C}_{0})\bigcap\mathcal{C}\setminus \mathcal{C}_{0}$.
\end{lemma}
{\bf Proof:} By Lemma \ref{le1}, $\mathcal{C}$ is a $2$-FPC. By the definition of an FPC, we have ${\sf desc}(\mathcal{C}_{0})\bigcap\mathcal{C}=\mathcal{C}_{0}$ when $|\mathcal{C}_{0}|=1$, $2$. This implies ${\sf desc}(\mathcal{C}_{0})\bigcap\mathcal{C}\setminus \mathcal{C}_{0}=\emptyset$. Clearly the statement holds. So we only need to consider the case $|\mathcal{C}_{0}|=3$. For any $\mathcal{C}_{0}=\{{\bf c}_1,{\bf c}_2,{\bf c}_3\}$, where ${\bf c}_i=(a_{i},b_{i},e_{i})^{T}$, $1\leq i\leq 3$, suppose that there exits one codeword
${\bf c}'=(a',b',e')^{T} \in {\sf desc}(\mathcal{C}_{0})\bigcap\mathcal{C}\setminus \mathcal{C}_{0}$, such that $d({\bf c }_1, {\bf c}')=1$. Without loss of generality,
assume $a_1=a'$, $b_1=b'$, $e_1\neq e'$. This implies that $e'$ equals $e_2$ or $e_3$ since ${\bf c}' \in {\sf desc}(\mathcal{C}_{0})$. If $e'=e_2$ (or $e'=e_3$), we have ${\bf c}' \in {\sf desc}(\{{\bf c}_1, {\bf c}_2\})$ (or ${\bf c}' \in {\sf desc}(\{{\bf c}_1, {\bf c}_3\})$),
a contradiction to the definition of a $2$-FPC. So the statement also holds when $|\mathcal{C}_{0}|=3$. \qed
\begin{theorem}\rm
\label{th1}
For any $q \geq 3$, an $(n,M,q)$ code $\mathcal{C}$ is a $\overline{3}$-SSC$(3,M,q)$ if and only if
$\mathcal{C}$ is a $\overline{3}$-SC$(3,M,q)$.
\end{theorem}
{\bf Proof:} The necessity of the condition directly follows from Lemma \ref{le3}.
We now show that any $\overline{3}$-SC$(3,M,q)$ $\mathcal{C}$ over $Q$ is also a $\overline{3}$-SSC$(3,M,q)$. That is, for any $\mathcal{C}_{0}\subseteq\mathcal{C}$, $|\mathcal{C}_{0}|\leq 3$,
we should show $\cap_{\mathcal{C}'\in\hbox{S}(\mathcal{C}_{0})}\mathcal{C}'=\mathcal{C}_{0}$ from the definition of an SSC.

By Lemma \ref{le1}, $\mathcal{C}$ is a $2$-FPC. From Lemma \ref{le2}, we have $\mathcal{C}$ is a $\overline{2}$-SSC.
So when $|\mathcal{C}_{0}|=1$, $2$, $\cap_{\mathcal{C}'\in\hbox{S}(\mathcal{C}_{0})}\mathcal{C}'=\mathcal{C}_{0}$ holds.
Now we consider the case $|\mathcal{C}_{0}|=3$. For any $\mathcal{C}_{0}=\{{\bf c}_{1},{\bf c}_{2},{\bf c}_{3}\}$,
${\bf c}_{i}=(a_{i},b_{i},e_{i})$, we have ${\sf desc}(\mathcal{C}_{0})$:
\begin{eqnarray}
\label{array1}
\left(
\begin{array}{ccccccccccccccccccccccccccc}
a_{1} & a_{2} & a_{3} & a_{1} & a_{2} & a_{1} & a_{2} & a_{3} & a_{3} & a_{1} & a_{1} & a_{1} & a_{1} & a_{1} & a_{1} & a_{2}& a_{2} & a_{2} & a_{2} & a_{2} & a_{2} & a_{3} & a_{3} & a_{3} & a_{3} & a_{3} & a_{3} \\
b_{1} & b_{2} & b_{3} & b_{2} & b_{1} & b_{3} & b_{3} & b_{1} & b_{2} & b_{1} & b_{1} & b_{2} & b_{2} & b_{3} & b_{3} & b_{2} & b_{3} & b_{3} & b_{2} & b_{1} & b_{1} & b_{3} & b_{3} & b_{1} & b_{1} & b_{2} & b_{2} \\
e_{1} & e_{2} & e_{3} & e_{3} & e_{3} & e_{2} & e_{1} & e_{2} & e_{1} & e_{2} & e_{3} & e_{1} & e_{2} & e_{1} & e_{3} & e_{3} & e_{2} & e_{3} & e_{1} & e_{2} & e_{1} & e_{1} & e_{2} & e_{3} & e_{1} & e_{3} & e_{2}
\end{array}
\right) \ \ \
\end{eqnarray}
Let ${\bf c}_{i}$, $1\leq i\leq27$, be the $i$th codeword of ${\sf desc}(\mathcal{C}_{0})$ in \eqref{array1}.

According to Lemma \ref{le4}, ${\bf c}_{i} \notin {\sf desc}(\mathcal{C}_{0})\bigcap\mathcal{C}$, $10 \leq i\leq 27$.
Hence we have
\begin{eqnarray}
\label{desc}
{\sf desc}(\mathcal{C}_{0})\bigcap \mathcal{C} \subseteq \left(
\begin{array}{ccccccccccccccccccccccccccc}
a_{1} & a_{2} & a_{3} & a_{1} & a_{2} & a_{1} & a_{2} & a_{3} & a_{3}\\
b_{1} & b_{2} & b_{3} & b_{2} & b_{1} & b_{3} & b_{3} & b_{1} & b_{2}\\
e_{1} & e_{2} & e_{3} & e_{3} & e_{3} & e_{2} & e_{1} & e_{2} & e_{1}
\end{array}
\right) \ \ \
\end{eqnarray}
Now we consider formula  \eqref{desc} by discussing cardinalities of sets $\mathcal{C}_{0}(i)$, $1\leq i \leq 3$.
\begin{itemize}
\item If there exists an integer $1 \leq i \leq 3$ such that $|\mathcal{C}_{0}(i)| \leq 2$, without loss of generality, assume $|\mathcal{C}_{0}(1)| \leq 2$ and $a_{1}=a_{2}$. According to Lemma \ref{le4}, we have ${\bf c}_{i} \notin {\sf desc}(\mathcal{C}_{0})\bigcap\mathcal{C}$, $4 \leq i\leq 7$. So we only need to consider ${\bf c}_8$ and ${\bf c}_9$.

\begin{itemize}
\item If $a_{1}=a_{3}$, then ${\bf c}_8, {\bf c}_9 \notin {\sf desc}(\mathcal{C}_{0})\bigcap\mathcal{C}$ from Lemma \ref{le4}.
So $\cap_{\mathcal{C}'\in\hbox{S}(\mathcal{C}_{0})}\mathcal{C}'=\mathcal{C}_{0}$.
\item If $a_{1}\neq a_{3}$, $|\{b_1,b_2,b_3\}|<3$ or $|\{e_1,e_2,e_3\}|<3$ holds,  then ${\bf c}_8, {\bf c}_9 \notin {\sf {\sf desc}}(\mathcal{C}_{0})\bigcap\mathcal{C}$ from Lemma \ref{le4}.
So $\cap_{\mathcal{C}'\in\hbox{S}(\mathcal{C}_{0})}\mathcal{C}'=\mathcal{C}_{0}$.
\item If $a_{1}\neq a_{3}$ and $|\{b_1,b_2,b_3\}|=|\{e_1,e_2,e_3\}|=3$, then ${\sf desc}(\mathcal{C}_{0})\bigcap\mathcal{C}$ contains at most one of ${\bf c}_8$ and ${\bf c}_9$. Otherwise, we have ${\sf desc}(\{{\bf c}_1,{\bf c}_2,{\bf c}_8\})=$ ${\sf desc}(\{{\bf c}_1$, ${\bf c}_2, {\bf c}_9\})$,
a contradiction to the definition of a $\overline{3}$-SC.
Without loss of generality, suppose that ${\bf c}_8 \in {\sf desc}(\mathcal{C}_{0})\bigcap\mathcal{C}$.
We have ${\sf desc}(\mathcal{C}_{0})\bigcap\mathcal{C}=\{{\bf c}_{1},{\bf c}_{2},{\bf c}_{3},{\bf c}_{8}\}$.
Clearly S$(\mathcal{C}_{0})=\{\mathcal{C}_{0},\mathcal{C}_{1}\}$ where $\mathcal{C}_{1}=\{{\bf c}_{1},{\bf c}_{2},{\bf c}_{3},{\bf c}_{8}\}$.
It is easy to check that $\cap_{\mathcal{C}' \in \hbox{S}(\mathcal{C}_{0})}\mathcal{C}'=\mathcal{C}_{0}$.
\end{itemize}
\item  If $|\mathcal{C}_{0}(i)|=3$ for any $1 \leq i \leq 3$, it is easy to check that $d({\bf c}_{j_1},{\bf c}_{j_2})=2$ for all $j_1=1,2,3$
and $j_2=4,5,\ldots, 9$. If ${\sf desc}(\mathcal{C}_{0})\bigcap\mathcal{C}=\mathcal{C}_{0}$,  clearly it is a $\overline{3}$-SSC. Now we consider the case that there is at least one codeword ${\bf c}_i \in {\sf desc}(\mathcal{C}_{0})\bigcap\mathcal{C}$,
$4\leq i\leq 9$, without loss of generality, we assume ${\bf c}_4 \in {\sf desc}(\mathcal{C}_{0})\bigcap\mathcal{C}$. By the distance, ${\bf c}_{i}$,
$5\leq i\leq 9$, can be divided into two subsets $\mathcal{C}_1=\{{\bf c}_{5},{\bf c}_{6},{\bf c}_{9}\}$ and $\mathcal{C}_2=\{{\bf c}_{7},{\bf c}_{8}\}$ such that $d({\bf c}_{4},{\bf c})=2$ if ${\bf c}\in \mathcal{C}_1$ and $d({\bf c}_{4},{\bf c})=3$ if ${\bf c}\in \mathcal{C}_2$.
\begin{itemize}
\item If ${\sf desc}(\mathcal{C}_{0})\cap\mathcal{C}=\{{\bf c}_{1},{\bf c}_{2},{\bf c}_{3},{\bf c}_{4}\}$,
then S$(\mathcal{C}_{0})=\{\mathcal{C}_{0},\mathcal{C}'\}$, where $\mathcal{C}'=\{{\bf c}_{1}, {\bf c}_{2},$ \\ $ {\bf c}_{3}, {\bf c}_{4}\}$.
Clearly $\cap_{\mathcal{C}' \in \hbox{S}(\mathcal{C}_{0})}\mathcal{C}'=\mathcal{C}_{0}$.
\item If $\{{\bf c}_{1},{\bf c}_{2},{\bf c}_{3},{\bf c}_{4}, {\bf c}\}\subseteq{\sf desc}(\mathcal{C}_{0})\cap\mathcal{C}$, ${\bf c}\in \mathcal{C}_1$, we claim that this case does not happen. We take ${\bf c}={\bf c}_5$ as an example. Then we have ${\sf desc}(\{{\bf c}_{1},{\bf c}_{2},{\bf c}_{4}\})=$\ ${\sf desc}(\{{\bf c}_{1},{\bf c}_{2},{\bf c}_{5}\})$, a contradiction to the definition of a $\overline{3}$-SC.
\item If ${\sf desc}(\mathcal{C}_{0})\cap\mathcal{C}=\{{\bf c}_{1},{\bf c}_{2},{\bf c}_{3},{\bf c}_{4}, {\bf c}\}$, ${\bf c}\in \mathcal{C}_2$, we claim that this case satisfies the conditions of $\overline{3}$-SSC. We take ${\bf c}={\bf c}_7$ as an example. Let $\mathcal{C}'=\{{\bf c}_{1},{\bf c}_{2},{\bf c}_{3},{\bf c}_{4}\}$, $\mathcal{C}''=\{{\bf c}_{1},{\bf c}_{2},{\bf c}_{3},{\bf c}_{7}\}$ and $\mathcal{C}'''=\{{\bf c}_{1},{\bf c}_{2},{\bf c}_{3},{\bf c}_{4},{\bf c}_{7}\}$. It is easy to check that $S(\mathcal{C}_{0})=\{\mathcal{C}_{0},\mathcal{C}',\mathcal{C}'',\mathcal{C}'''\}$ and $\cap_{\mathcal{C}' \in \hbox{S}(\mathcal{C}_{0})}\mathcal{C}'=\mathcal{C}_{0}$.
\item If ${\sf desc}(\mathcal{C}_{0})\cap\mathcal{C}=\{{\bf c}_{1},{\bf c}_{2},{\bf c}_{3},{\bf c}_{4}, {\bf c}_7,{\bf c}_8\}$, then ${\sf desc}(\{{\bf c}_{1},{\bf c}_{2},{\bf c}_{3}\})= {\sf desc}(\{{\bf c}_{4},{\bf c}_{7},$ \\ ${\bf c}_{8}\})$, a contradiction to the definition of a $\overline{3}$-SC.
\end{itemize}
\end{itemize}
From the above discussions, we know that $\mathcal{C}$ is a $\overline{3}$-SSC. Then the proof is complete.\qed

\subsection{Upper bounds on $M(\overline{t},n,q)$}\

As an important class of anti-collusion codes in multimedia copyright protection, separable codes and frameproof codes were widely studied, e.g.,
\cite{B4,C.F.J.L.M,C.J.M,C.J.L.M.T,C.J.T,C.M,G.G}.

\begin{theorem} \rm(\cite{C.J.M})
 \label{th2}
Given a $\overline{t}$-SC$(n,M,q)$ with $t \ge 3$ and $n \ge 2$, let $r \in \{0,1,\ldots,t-2\}$ be the remainder of $n$ on division by $t-1$.
If $M > q$, then
$$ M \le {\max}\{q^{\lceil n/(t-1) \rceil}, r(q^{\lceil n/(t-1) \rceil}-1)+(t-1-r)(q^{\lfloor n/(t-1) \rfloor}-1)\}.$$
\end{theorem}

\begin{lemma}\rm (\cite{B2})
\label{le5}
In a $\overline{2}$-SC$(n,M,q)$, we have $M \leq q^{\lceil\frac{2n}{3}\rceil} + \frac{1}{2}q^{\lfloor\frac{n}{3}\rfloor}(q^{\lfloor\frac{n}{3}\rfloor} -1)$.
\end{lemma}

\begin{theorem}\rm (\cite{C.F.J.L.M})
\label{th3}
For any $\overline{2}$-SC$(2,M,q)$, we have $M \leq qk+t$,
where $k=\lfloor \frac{1+\sqrt{4q-3}}2 \rfloor$, and
\[  t = \left\{
    \begin{array}{ll}
       0 & {\it if~~} k^2-k+1\leq q\leq k^2-1;\\
       \lfloor \frac{(3k^2+k-1)-\sqrt{5k^4+6k^3-k^2-2k+1}}2 \rfloor & {\it if~~} q = k^2;\\
       \lfloor \frac{(k-1)q}{(k+1)^2-(q+1)}\rfloor & {\it if~~} k^2+1\leq q\leq k^2+k-2;\\
       k^2-k & {\it if~~} q = k^2+k-1;\\
       k^2 & {\it if~~} q = k^2+k.
 \end{array} \right. \]
Furthermore, there always exists an optimal $\overline{2}$-SC$(2,M,q)$ if $q \in \{ k^2 -1, k^2+k-2, k^2+k-1, k^2+k, k^2+k+1\}$ for any prime power $k\geq 2$.
\end{theorem}
\begin{lemma} \rm(\cite{C.J.L.M.T})
\label{le6}
For any positive integers $t$ and $q\geq 2$.
\begin{itemize}
\item When $2\leq n< t$, there always exists an optimal $\overline{t}$-SC$(n,n(q-1),q)$ and an optimal $t$-FPC$(n,n(q-1),q)$.
\item When $n=t$, for any $\overline{t}$-SC$(n,M,q)$ we have $M\leq q^2$ if $n\leq q$, otherwise $M\leq nq$.
\end{itemize}
\end{lemma}

When $t=3$, $n=3$, Cheng et al. improved the upper bound in Lemma \ref{le6}.
\begin{lemma} \rm (\cite{C.J.L.M.T})
\label{le7}
In a $\overline{3}$-SC$(3,M,q)$ with $q\geq4$, we have $M\leq\lfloor\dfrac{3q^2}{4}\rfloor$.
\end{lemma}
By the constructions of PHFs, Cheng et al. also proposed a lower bound.
\begin{lemma} \rm (\cite{C.J.L.M.T})
\label{le8}
There always exists a $\overline{3}$-SC$(3,M,q)$ with $M\geq\lfloor\sqrt{q}\rfloor^3$.
\end{lemma}
From the above results and Lemma \ref{le3}, the following upper bounds on $M(\overline{t},n,q)$ can be obtained.
\begin{theorem} \rm
\label{th4}
Let $n$, $q$ and $t$ be positive integers such that $t \ge 3$ and $n \ge 2$,
and let $r \in \{0,1,\ldots,t-2\}$ be the remainder of $n$ on division by $t-1$.
If $M(\overline{t},n,q) > q$, then
$$ M(\overline{t},n,q) \le {\max}\{q^{\lceil n/(t-1) \rceil}, r(q^{\lceil n/(t-1) \rceil}-1)+(t-1-r)(q^{\lfloor n/(t-1) \rfloor}-1)\}.$$
\end{theorem}
\begin{lemma}\rm
\label{le9}
$M(\overline{2},n,q) \leq q^{\lceil\frac{2n}{3}\rceil} + \frac{1}{2}q^{\lfloor\frac{n}{3}\rfloor}(q^{\lfloor\frac{n}{3}\rfloor} -1)$.
\end{lemma}
\begin{theorem}\rm
\label{th5}
For any positive integer $q$, $M(\overline{2},2,q) \leq qk+t$,
where $k=\lfloor \frac{1+\sqrt{4q-3}}2 \rfloor$, and
\[  t = \left\{
    \begin{array}{ll}
       0 & {\it if~~} k^2-k+1\leq q\leq k^2-1;\\
       \lfloor \frac{(3k^2+k-1)-\sqrt{5k^4+6k^3-k^2-2k+1}}2 \rfloor & {\it if~~} q = k^2;\\
       \lfloor \frac{(k-1)q}{(k+1)^2-(q+1)}\rfloor & {\it if~~} k^2+1\leq q\leq k^2+k-2;\\
       k^2-k & {\it if~~} q = k^2+k-1;\\
       k^2 & {\it if~~} q = k^2+k.
 \end{array} \right. \]
Furthermore, $M(\overline{2},2,q) = qk+t$ if $q \in \{ k^2 -1, k^2+k-2, k^2+k-1, k^2+k, k^2+k+1\}$ for any prime power $k\geq 2$.
\end{theorem}

From Lemmas \ref{le2}, \ref{le3} and \ref{le6}, the following statement holds.
\begin{lemma} \rm
\label{le10}
For any positive integers $t$ and $q\geq 2$.
\begin{itemize}
\item When $2\leq n< t$, $M(\overline{t},n,q)=n(q-1)$.
\item When $n=t$, if $M(\overline{t},n,q)\leq q^2$, and otherwise $M(\overline{t},n,q)\leq nq$.
\end{itemize}
\end{lemma}

From Theorem \ref{th1} and Lemmas \ref{le7}, \ref{le8}, we have the following result.
\begin{theorem} \rm
\label{th6}
$\lfloor\sqrt{q}\rfloor^3\leq M(\overline{3},3,q)\leq\lfloor\dfrac{3q^2}{4}\rfloor$ holds for $q\geq4$.
\end{theorem}

To our best knowledge, the lower bound in \cite{C.J.L.M.T} is the best known result on $\overline{3}$-SC$(3,M,q)$. In the following section, we will improve the lower bound in Theorem \ref{th6} to $\Omega (q^{5/3}+q^{4/3}-q)$ for some prime powers $q$.

\section{Construction}
\label{se-construction}
From  Theorem \ref{th1}, it is sufficient to consider $\overline{3}$-SC$(3,M,q)$ for studying $\overline{3}$-SSC$(3,M,q)$. First the following notations are necessary.

For any $(3,M,q)$ code $\mathcal{C}$ defined on $Q=\{0, 1, \cdots, q-1\}$,
we define the column vector sets $\mathcal{A}_i^{(1)}$ for $i \in Q$ as follows:
\begin{eqnarray*}
\label{aij}
\mathcal{A}_i^{(1)} = \{(x_2, x_3)^T \ | \ (x_1, x_2, x_3)^{T} \in \mathcal{C}, \ x_1=i\}.
\end{eqnarray*}
Obviously, $\mathcal{A}_i^{(1)} \subseteq Q^{2}$ for any
$i \in Q$ and $|\mathcal{A}_0^{(1)}|+ \cdots+ |\mathcal{A}_{q-1}^{(1)}|=M$
hold. Similar to the above notation, vector sets ${\mathcal{A}}_i^{(j)}$ for $j=2$, $3$  can be also defined.
\begin{lemma}\rm(\cite{C.J.L.M.T})
\label{le11}
A $(3,M,q)$ code is a $2$-FPC$(3,M,q)$ if and only if $|{\mathcal{A}}_{i}^{(j)} \bigcap {\mathcal{A}}_{i'}^{(j)}|\leq 1$ holds
for any $j\in \{1,2,3\}$ and distinct $i$, $i'\in Q$, where if $|\mathcal{A}_{i}^{(j)} \bigcap \mathcal{A}_{i'}^{(j)}|= 1$, then $|\mathcal{A}_{i}^{(j)}|=|\mathcal{A}_{i'}^{(j)}|= 1$.
\end{lemma}

Cheng et al. showed that for any $\overline{3}$-SC$(3,M,q)$, $\mathcal{C}$, there is no subcode $\triangle_{i}\subseteq \mathcal{C}$ described in \eqref{forb1}, $a\neq b$, $c\neq d$, $e\not\in\{f,g\}$, and there is no subcode $\nabla\subseteq \mathcal{C}$ described in \eqref{forb2}, $|\{a_i,b_i,c_i\}|=3$, $i=1$, $2$, $3$.
\begin{eqnarray}
\label{forb1}
&\triangle_{1}=\left(
\begin{array}{cccc}
a & a & b & b \\
e & f & g & e \\
c & d & c & d
\end{array}
\right)\ \ \
\triangle_{2}=\left(
\begin{array}{cccc}
a & a & b & b \\
c & d & c & d \\
e & f & g & e
\end{array}
\right)
&\triangle_{3}=\left(
\begin{array}{cccc}
e & f & g & e \\
a & a & b & b \\
c & d & c & d
\end{array}
\right)
\end{eqnarray}
\begin{eqnarray}
\label{forb2}\nabla=\left(
\begin{array}{cccccc}
a_{1} & b_{1}  &c_{1} & a_{1}&b_{1} & c_{1}\\
a_{2} & b_{2} &c_{2} & b_{2}&c_{2} & a_{2}\\
a_{3} & b_{3} &c_{3} & c_{3}&a_{3} & b_{3}
\end{array}
\right)
\end{eqnarray}

We call such $\triangle_{i}$ and $\triangle$  {\em forbidden configurations} of $\mathcal{C}$.
\begin{theorem} \rm(\cite{C.J.L.M.T})
\label{th7}
A $(3,M,q)$ code $\mathcal{C}$ is a $\overline{3}$-SC$(3,M,q)$ if and only if it satisfies the following conditions:
\begin{itemize}
\item[(i)] $\mathcal{C}$ is a $2$-FPC$(3,M,q)$;
\item[(ii)]  Configurations in \eqref{forb1} and \eqref{forb2} are all the forbidden configurations of $\mathcal{C}$.
\end{itemize}
\end{theorem}

In the following, for any prime power $q$, we will take advantage of difference matrix to construct $\overline{3}$-SC$(3,M,q)$.
\begin{definition} \rm(\cite{C.M})
\label{de2}
For any prime power $q$, a \emph{difference matrix} $(q,3,1)$DM is a $3 \times q$ matrix $D = (d_{j, i})$ with $d_{j, i} \in \mathbb{F}_q$
such that for any $1 \le s \ne t \le 3$, the differences $d_{s,i}- d_{t,i}$ over $\mathbb{F}_q$, $i\in \mathbb{F}_q$, comprise all the elements of $\mathbb{F}_q$.
\end{definition}
Given a $3 \times s$ matrix $N$ with entries from $\mathbb{F}_q$ and $s$ distinct columns ${\bf n}_1, {\bf n}_2, \ldots, {\bf n}_s$,
we can define a $(3,qs,q)$ code $\mathcal{C}$ on $\mathbb{F}_q$ as
$$\mathcal{C}=\{N+g \ | \ g \in \mathbb{F}_q\} = \{{\bf n}_i+g \ | \ g \in \mathbb{F}_q, \ 1 \le i \le s\}.$$
We say $N$ is a base of $\mathcal{C}$, or $\mathcal{C}$ is generated by $N$.

For any given $(q,3,1)$DM, $D$, we can obtain a $(3,q^2,q)$ code $\mathcal{C}=\{D+g \ | \ g \in \mathbb{F}_q\}$.
By the definition of a DM, we know that $|\mathcal{A}_{i_1}^{(j)} \cap \mathcal{A}_{i_2}^{(j)}|=0$ holds
for any $1 \le j \le 3$ and for any distinct $i_1 ,i_2 \in \mathbb{F}_q $, which implies that
$\mathcal{C}$ is a $2$-FPC$(3,q^2,q)$ by Lemma \ref{le11}.
Unfortunately, this code is not always a $\overline{3}$-SC.
\begin{example} \rm
The following code $\mathcal{C}$ is generated by $(3,3,1)$DM. Let ${\bf c}_{i}$ denote the $i$th codewode, $1\leq i\leq9$. From above discussion, $\mathcal{C}$ is a $2$-FPC$(3,9,3)$, but is not a $\overline{3}$-SC since ${\sf desc}(\{{\bf c}_{1},{\bf c}_{4},{\bf c}_{7}\}) = {\sf desc}(\{{\bf c}_{2},{\bf c}_{5},{\bf c}_{8}\})$.
\begin{eqnarray*}
\mathcal{C}=\left(
\begin{array}{ccccccccc}
0 & 0 & 0 & 1 & 1 & 1 & 2 & 2 & 2 \\
0 & 1 & 2 & 1 & 2 & 0 & 2 & 0 & 1\\
0 & 2 & 1 & 1 & 0 & 2 & 2 & 1 & 0
\end{array}
\right)
\end{eqnarray*}
\end{example}

In fact, we can obtain the base of a $\overline{3}-$SC$(3,M,q)$ by deleting some codewords in $\mathcal{C}$ generated by $(q,3,1)$DM. For any prime power $q$, if $q\geq 3$, it is easy to check that the following array $D$ is a $(q,3,1)$DM
\begin{eqnarray}
\label{fo12}
D=\left(
\begin{array}{cccc}
0 & 0 & ... & 0 \\
0 & 1 & ...& \varepsilon^{q-2} \\
0 & \alpha & ... & \alpha\varepsilon^{q-2}
\end{array}
\right),
\end{eqnarray}
where $\varepsilon$ is a primitive element of $\mathbb{F}_q$ and $\alpha$ is an element of  $\mathbb{F}_q\setminus \{0,1\}$.
For any subset $S\subseteq \mathbb{F}_q$,  let sub-matrix $N=D|_S$ obtained by deleting the columns $i\in \mathbb{F}_q\setminus S$.
Clearly the code $\mathcal{C}$ generated by $N$ is a $2$-FPC$(3,q|S|,q)$.
From Theorem \ref{th7}, in order that $\mathcal{C}$ may be a $\overline{3}$-SC$(3,M,q)$, we only need to consider the forbidden configurations in \eqref{forb1} and \eqref{forb2}.
Suppose $C \subseteq\mathcal{C}$,
\begin{itemize}
\item[(I)]
When $C \in \{\triangle_1, \triangle_2, \triangle_3\}$, we may assume
\begin{eqnarray*}
C=\left(\begin{array}{cccc}
k_1& k_2&k_3&k_4 \\
x+k_1& y+k_2&z+k_3&w+k_4 \\
\alpha x+k_1&\alpha y+k_2&\alpha z+k_3&\alpha w+k_4 \ \ \
\end{array}\right),
\end{eqnarray*}
where $x,y,z,w,k_1,k_2,k_3,k_4\in \mathbb{F}_q$.
\begin{itemize}
\item When $C =\triangle_1$, we have
$$k_1 =k_2, \ x+k_1=w+k_4, \ \alpha y+k_2=\alpha w +k_4.$$

This means
\begin{eqnarray}
\label{equ0}
x+(\alpha-1)w=y\alpha
\end{eqnarray}
with $|\{x,y,w\}|=3$. In fact, if $x=y$, we have that the first codeword equals the second codeword of $\triangle_1$, a contradiction to the assumption. This implies $x\neq y$. Similarly we can check that $x\neq w$ and $y\neq w$.
\item When $C =\triangle_2$, we have
$$k_1 =k_2,\ y+k_2=w+k_4, \  \alpha x+k_1=\alpha w +k_4.$$
This means
\begin{eqnarray}
\label{equ1}
y+(\alpha-1)w=x\alpha.
\end{eqnarray}
It is easy to check that  $|\{x,y,w\}|=3$ holds in \eqref{equ1}.
\item When $C =\triangle_3$, we have
$$k_1 =k_4,\  x+k_1=y+k_2, \ \alpha y +k_2=\alpha w +k_4.$$
This means
\begin{eqnarray}
\label{equ2}
x+(\alpha-1)y=w\alpha.
\end{eqnarray}
It is easy to check that $|\{x,y,w\}|=3$ holds in \eqref{equ2}.
\end{itemize}

\item[(II)]
 When $C =\nabla$, we may assume
\begin{eqnarray*}
C =\left(
 \begin{array}{cccccc}
 k_1          & k_2         &k_3        & k_1         &k_2         & k_3\\
 x+k_1       & y+k_2       &z+k_3       &u+ k_1       &v+k_2       &w+ k_3\\
 \alpha x+k_1& \alpha y+k_2&\alpha z+k_3&\alpha u+ k_1&\alpha v+k_2&\alpha w+ k_3
 \end{array}
 \right) .\ \ \
\end{eqnarray*}
\begin{eqnarray}
\label{equ3-1}
\left\{\begin{array}{ll}
x+k_1=w+k_3\\
y+k_2=u+k_1\\
z+k_3=v+k_2\\
\alpha x+ k_1=\alpha v+ k_2\\
\alpha y+ k_2=\alpha w+ k_3\\
\alpha z+ k_3=\alpha u+ k_1
\end{array}
\right.\Longrightarrow
\left\{\begin{array}{lllll}
k_3-k_1=x-w\\
k_2-k_1=u-y\\
k_3-k_2=v-z\\
k_2-k_1=\alpha x-\alpha v\\
k_3-k_2=\alpha y-\alpha w\\
k_3-k_1=\alpha u-\alpha z
\end{array}.
\right.
\end{eqnarray}

This means
\begin{eqnarray}
\label{abc4}
\left\{\begin{array}{ll}
\alpha x +\alpha(\alpha-1)z=(\alpha-1)y+(\alpha^2-\alpha+1)u\\
\alpha w +\alpha(\alpha-1)u=(\alpha-1)v+(\alpha^2-\alpha+1)z
\end{array}
\right.
\end{eqnarray}
Then we know $\{x,y,z\}\cap \{u,v,w\}=\emptyset$ always holds.
\begin{itemize}
\item $x\not\in\{u,v,w\}$ always holds. If $x=u$, we have
the first codeword  equals the forth codeword of $\nabla$, a contradiction to the assumption.
Similarly, we can prove that $x \neq w$, $v$ always holds.
\item  $y\not\in\{u,v,w\}$ always holds. If $y=u$, we have $k_1=k_2$ from $y+k_2=u+k_1$.
This implies that the second codeword  equals the forth codeword of $\nabla$, a contradiction to the assumption.
Similarly, we can prove that $y \neq w$, $v$.
\item $z\not\in\{u,v,w\}$ always holds. If $z=u$, we have $k_1=k_3$ from
 $\alpha z+k_3=\alpha u+k_1$.
This implies that the third codeword equals the forth codeword of  $\nabla$, a contradiction to the assumption.
Similarly, we have $z \neq w$, $v$.
\end{itemize}
\end{itemize}

For any prime power $q_1$ and positive integer $n$, let $\mathbb{F}^n_{q_1}$ be the $n$-dimensional vector space over  $\mathbb{F}^n_{q_1}$. When $q=q_1^n$, it is well known that the element of $\mathbb{F}_q$ can be represented by the $n$-dimensional vector over $\mathbb{F}_{q_1}$. Suppose that $S$ is a subset of $\mathbb{F}^n_{q_1}$, of which no three distinct elements are collinear. Then equations \eqref{equ0}, \eqref{equ1} and \eqref{equ2} have no solution in $S$.
This implies that $\mathcal{C}$ does not contain $\triangle_1$, $\triangle_2$ and $\triangle_3$.
Together with \eqref{abc4}, we have the following result.
\begin{theorem}\rm
\label{th8}
For any subset $S\subseteq \mathbb{F}_q$, of which no three distinct elements are collinear, if there is no solution of \eqref{abc4} in $S$, then the code generated by $N=D|_S$ is a $\overline{3}$-SC$(3,q|S|,q)$.
\end{theorem}

Now, we focus on the formula \eqref{abc4}. Let $q_1=6t+1$ be a prime power, and $\alpha$ be a primitive $6$th root of unity in $\mathbb{F}_{q_1}$, where $t\geq 1$. Clearly $\alpha$ is a root of $f(x)=x^2-x+1$. Then \eqref{abc4} can be written as
\begin{eqnarray}
\label{equ7-0}
\left\{\begin{array}{ll}
x +(\alpha-1)z=\alpha y \\
w +(\alpha-1)u=\alpha v
\end{array}
\right.
\end{eqnarray}

From \eqref{equ7-0}, if $|\{x,y,z\}|<3$ (or $|\{u,v,w\}|<3$), then $|\{x,y,z\}|=1$ (or $|\{u,v,w\}|$ \\ $=1$) always holds. Furthermore, from \eqref{equ3-1} we claim if $x=y=z$ (or $u=v=w$), then $|\{u,v,w\}|=3$ (or $|\{x,y,z\}|=3$) always holds in \eqref{abc4}. If $x=y=z$ and $u=v=w$, then $x+k_1=w+k_3$ and $\alpha x+k_3=\alpha w+k_1$ hold by \eqref{equ3-1}. We have $(\alpha+1)x=(\alpha+1)w$. This implies $x=w$, $k_1=k_3$ since $\alpha\neq -1$. That is, the first codeword equals the sixth codeword in $C$, a contradiction. So we have
\begin{eqnarray}
\label{equ6-6}
\begin{array}{lllll}
|\{x,y,z,u,v,w\}|=6; or\\
|\{x,y,z\}|=3\ \ \hbox{and} \ \ |\{u,v,w\}|=1; or\\
|\{x,y,z\}|=1\ \ \hbox{and} \ \ |\{u,v,w\}|=3.
\end{array}
\end{eqnarray}

According to \eqref{equ7-0} and \eqref{equ6-6}, Theorem \ref{th8} can be written as follows.
\begin{theorem}\rm
\label{th9}
Let $q=q_{1}^n$, where $q_1=6t+1$ is a prime power, $t\geq 1$.
For any subset $S\subseteq \mathbb{F}_q$, of which no three distinct elements are collinear,
the code generated by $N=D|_S$ is a $\overline{3}$-SC$(3,q|S|,q)$.
\end{theorem}

Denoting by $r(q_{1}^n)$ the maximum size of a subset of $\mathbb{F}^n_{q_1}$ that contains no three points on a
line. There are many studies on the value of $r(q_{1}^n)$ over $\mathbb{F}^n_{q_1}$.
The interested reader is referred to \cite{L.C,Y.J,E.S}.
\begin{lemma} \rm(\cite{Y.J})
\label{le12}
For any prime power $q_1\geq 3$, we have $r(\mathbb{F}^{6}_{q_1})=\Omega(q_{1}^{4}+q_{1}^{2}-1)$.
\end{lemma}
From Theorem \ref{th9} and Lemma \ref{le12}, the following lower bound can be obtained.
\begin{lemma} \rm
\label{le13}
For any prime power $q=q^{6}_1$, where $q_{1}=6t+1$ is a prime power, there exists a $\overline{3}$-SSC$(3,M,q)$, where $M=\Omega (q^{5/3}+q^{4/3}-q)$.
\end{lemma}


\section{Conclusion} %
\label{Conclusion}       %

In this paper, we first derived several upper bounds on the number of codewords of $\overline{t}$-SSC.
Then we focused on $\overline{3}$-SSC with codeword length $3$, and obtained the following two main results:
(1) An equivalence between an SSC and an SC.
(2) An improved lower bound $\Omega (q^{5/3}+q^{4/3}-q)$ on the size of a $q$-ary SSC when $q=q_1^6$ for any prime power $q_1\equiv\ 1 \pmod 6$, better than the before known bound $\lfloor\sqrt{q}\rfloor^{3}$, which was obtained by means of difference matrix and the known result on the subset of $\mathbb{F}^{n}_q$ containing no three points on a line.

It would be of interest if we could improve the upper bounds 
$\lfloor\dfrac{3q^2}{4}\rfloor$ or the lower bound $\Omega (q^{5/3}+q^{4/3}-q)$. It would
be also interesting if we could get more properties and
constructions of strongly separable codes.
\section*{Acknowledgements}
This work is in supported by 2016GXNSF( No. FA380009 and CA380021) and 2014GXNSFDA(No. 118001).
\bibliographystyle{model1b-num-names}
\bibliography{<your-bib-database>}

\end{document}